\documentclass[aps,prd,twocolumn,nofootinbib]{revtex4-2}
\usepackage{graphicx}
\usepackage{mathrsfs}
\usepackage{color}
\graphicspath{{figures/}{fig/}}
\usepackage{amsmath}
\usepackage{amssymb}
\usepackage{bm}
\usepackage{slashed}
\usepackage{epsfig}
\usepackage{amsfonts}
\usepackage{epstopdf}
\usepackage{hyperref}
\usepackage{bbm}
\usepackage{textcomp}
\usepackage{color}
\usepackage[ruled,linesnumbered]{algorithm2e}

\newcommand{\sect}[1]{\section{#1}}
\newcommand{\mi}{\mathrm{i}}

\allowdisplaybreaks[3] 
\begin{document}

\title{Quark correlation functions at three-loop order and extraction of splitting functions}

\author{Chen Cheng}
\email{1801214966@pku.edu.cn}
\affiliation{School of Physics, Peking University, Beijing 100871, China}

\author{Li-Hong Huang}
\email{lhhuang@pku.edu.cn}
\affiliation{School of Physics, Peking University, Beijing 100871, China}

\author{Xiang Li}
\email{lix-PHY@pku.edu.cn}
\affiliation{School of Physics, Peking University, Beijing 100871, China}

\author{Zheng-Yang Li}
\email{zyli@jlab.org}
\affiliation{Theory Center, Jefferson Lab, 12000 Jefferson Avenue, Newport News, VA 23606, USA}

\author{Yan-Qing Ma}
\email{yqma@pku.edu.cn}
\affiliation{School of Physics, Peking University, Beijing 100871, China}
\affiliation{Center for High Energy Physics, Peking University, Beijing 100871, China}

\date{\today}

\begin{abstract}
We present the first complete next-to-next-to-next-to-leading-order calculation of the matching coefficients that link unpolarized flavor non-singlet parton distribution functions with lattice QCD computable correlation functions. By using this high-order result, we notice a reduction in theoretical uncertainties compared to relying solely on previously known lower-order matching coefficients. Furthermore, based on this result we have extracted the three-loop unpolarized flavor non-singlet splitting function, which is in agreement with the state-of-the-art result. Due to the simplicity of our method, it has the potential to advance the calculation of splitting functions to the desired four-loop order.
\end{abstract}

\maketitle

\sect{Introduction}
\label{sec:intro}
Parton distribution functions (PDFs) are important nonperturbative quantities that describe the one-dimensional internal structure of colliding hadrons. Since PDFs obey the Dokshitzer-Gribov-Lipatov-Altarelli-Parisi (DGLAP) evolution equations \cite{Gribov:1972ri,Altarelli:1977zs,Dokshitzer:1977sg}, whose kernel known as the splitting function can be computed perturbatively, it is sufficient to determine the PDFs at a specific renormalization scale. Presently, PDFs have been extracted from high-precision collider data \cite{Accardi:2016qay,Alekhin:2017kpj,Hou:2019efy,Bailey:2020ooq,NNPDF:2017mvq,NNPDF:2021njg,ATLAS:2021vod,Sitiwaldi:2023jjp}. To further enhance the accuracy of PDFs, it is crucial to compute the splitting functions to next-to-next-to-next-to-leading order (N3LO), i.e., four-loop order. Nevertheless, due to their complexity, the N3LO results are only known approximately so far \cite{Moch:2017uml, Falcioni:2023vqq, Falcioni:2023luc, Moch:2023tdj, Moch:2021qrk, Falcioni:2023tzp, Gehrmann:2023cqm, Gehrmann:2023iah}.

In addition to being extracted from experimental data, the first-principle lattice QCD (LQCD) calculations of PDFs are also extremely significant. Over the past decade, stimulated by the large momentum effective field theory~\cite{Ji:2013dva,Ji:2014gla}, new approaches for investigating PDFs through LQCD calculations have been developed and have led to tremendous breakthroughs in this field~\cite{Constantinou:2022yye}. The new methods include quantities such as quasi-PDFs~\cite{Ji:2013dva}, pseudo-PDFs~\cite{Radyushkin:2017cyf}, current-current correlators in momentum space~\cite{Chambers:2017dov}, and current-current correlators in position space~\cite{Ma:2017pxb}. These quantities can also be interpreted as good lattice cross sections~\cite{Ma:2014jla,Ma:2017pxb}, which are calculable using LQCD and factorizable to the universal PDFs, allowing us to extract PDFs from LQCD computations. Therefore, when using these methods, in addition to LQCD computations, it is crucial to compute the matching coefficients of factorization relations in perturbative QCD.

For the extraction of unpolarized flavor non-singlet PDFs, perturbative calculations have been carried out up to the next-to-next-to-leading order (NNLO) ~\cite{Chen:2020ody,Li:2020xml}. Based on these calculations, the unpolarized valence PDFs of both the pion and the proton have been obtained at the NNLO level through the analysis of LQCD computations~\cite{Gao:2021dbh,Bhat:2022zrw,Gao:2022iex,Gao:2022uhg}. These studies suggest that, for the extraction of PDFs, the NNLO corrections are generally smaller than the NLO corrections, indicating good perturbative convergence, which is crucial for precise calculations. Moreover, it was found that the NNLO results can lead to a reduction in scale-variation uncertainties compared to relying solely on NLO results. The aforementioned characteristics, along with the necessity of achieving high-precision PDFs, highlight the significance of higher-order perturbative calculations.

In this paper, we derive for the first time the N3LO non-singlet matching coefficients of the quark correlation function. By applying the N3LO matching coefficients to the valence-quark correlation functions, we show an improvement in perturbative uncertainty.
Furthermore, based on the obtained divergent part of the quark correlation function, we successfully extract the three-loop (or NNLO) unpolarized flavor non-singlet splitting function, which is in agreement with that obtained in the literature. Our method is much more simpler than traditional method using deep-inelastic structure functions \cite{Moch:2004pa}, and it may provide a feasible way to advance the calculation of the splitting functions to the desired N3LO.

\sect{Quark correlation functions}
Our study centers on the unpolarized gauge invariant quark correlation operator
\begin{align}
{\cal O}^{\nu,b}_q (\xi, \mu^2; \delta) = \Bar{\psi}^b_q (\xi) \gamma^\nu \Phi^{(f)} (\xi, 0) \psi^b_q (0) |_{\mu^2, \delta},
\end{align}
which has been utilized to define quasi-PDFs in Ref. \cite{Ji:2013dva}. The operator is composed of bare fields with a path ordered gauge link in fundamental representation, $\Phi^{(f)} (\xi, 0) = {\cal P} e^{-i g_s \int_0^1 dr \bold{\xi} \cdot \bold{A}^{(f)} (r\xi)}$. Here, $\delta$ is the ultraviolet (UV) regulator, $\mu$ is the UV renormalization scale, and $\bold{A}^{(f)} \equiv T^a \bold{A}^a$ is the Lie-algebra-valued gauge field in fundamental representation.
The UV divergences of the operator are known to be multiplicatively renormalizable \cite{Ji:2017oey,Ishikawa:2017faj,Green:2017xeu} as
\begin{align} \label{eq:OR}
{\cal O}^{\nu,\text{RS}}_q (\xi, \mu^2) = {\cal O}^{\nu,b}_q (\xi, \mu^2;\delta) / Z^{\text{RS}} (\xi^2, \mu^2; \delta),
\end{align}
where the superscript RS indicates a renormalization scheme and the renormalized operator is insensitive to the regulator $\delta$.

Quark correlation functions (QCFs) are defined as the hadronic matrix elements of ${\cal O}^{\nu,\text{RS}}_q$,
\begin{align}
F^{\nu,\text{RS}}_{q/h} (\omega,\xi^2,\mu^2) = \left \langle h(p) \right | {\cal O}^{\nu,\text{RS}}_q (\xi,\mu^2) \left | h(p) \right \rangle,
\end{align}
where $| h(p) \rangle$ is the state of a hadron $h$ with momentum $p$ and $\omega\equiv p\cdot \xi$. With $\xi_0 = 0$ and $\xi^2 \Lambda^2_{QCD} \ll 1$, $F^{\nu,\text{RS}}_{q/h}$ are calculable on LQCD and factorizable into PDFs \cite{Ma:2017pxb, Ma:2014jla, Izubuchi:2018srq}. Without loss of generality, we will assume that $\xi$ has only z-component in the rest of the paper.
In this paper, we focus on flavor non-singlet QCFs defined as 
\begin{equation}
   \begin{aligned} F^{\nu,{\text{RS}}}_{q_{ik}/h}(\omega, \xi^2, \mu^2)\equiv 
&F^{\nu,{\text{RS}}}_{q_{i}/h}(\omega, \xi^2, \mu^2)- F^{\nu,{\text{RS}}}_{q_{k}/h}(\omega, \xi^2, \mu^2).
\end{aligned}
\end{equation}
As any renormalization scheme RS can be related to the $\overline{\text{MS}}$ scheme as
\begin{align}
    F^{\nu,\text{RS}}_{q_{ik}/h} (\omega, \xi^2, \mu^2) = F^{\nu,\overline{\text{MS}}}_{q_{ik}/h} (\omega, \xi^2, \mu^2) / R^{\text{RS}}(\xi^2, \mu^2), 
\end{align}
where $R^{\text{RS}}(\xi^2, \mu^2) = Z^{\text{RS}}(\xi^2, \mu^2;\delta)/Z^{\overline{\text{MS}}}(\xi^2, \mu^2;\delta)$ is the finite renormalization factor, we will only focus on the $\overline{\text{MS}}$ renormalization scheme.

The flavor non-singlet QCFs in the $\overline{\text{MS}}$ scheme possess the following collinear factorization formula:
\begin{equation}
\begin{aligned} \label{eq:factorization}
&F^{\nu,\overline{\text{MS}}}_{q_{ik}/h} (\omega, \xi^2, \mu^2)= \int_{-1}^1 \frac{dx}{x} f_{q_{ik}/h}(x,\mu^2) K^{\nu}\left(x\omega, \xi^2, \mu^2\right) , 
\end{aligned}
\end{equation}
with corrections at $O(\xi^2 \Lambda^2_{QCD})$, where we have not differentiated the renormalization scale from the factorization scale. $f_{q_{ik}/h}(x,\mu^2) \equiv f_{q_{i}/h}(x,\mu^2) - f_{q_{k}/h}(x,\mu^2)$ is the flavor non-singlet PDF renormalized in $\overline{\text{MS}}$ scheme, and $K^{\nu}$ are perturbative matching coefficients which will be computed to N3LO in this paper.

As $Z^{\overline{\text{MS}}}(\xi^2, \mu^2;\delta)$ is the renormalization factor of the quark correlation operator and is independent of the property of the hadron $h$ and its momentum, it can be obtained by
\begin{align}
    \left. Z^{\overline{\text{MS}}}(\xi^2, \mu^2;\delta) = \frac{\tilde{p}_\nu F^{\nu,b}_{q_{ik}/q_i}(\omega, \xi^2, \mu^2;\delta)}{\tilde{p}_\nu F^{\nu,\overline{\text{MS}}}_{q_{ik}/q_i}(\omega, \xi^2, \mu^2)} \right|_{\omega \rightarrow 0},
\end{align}
where $\tilde{p}_\nu = \xi_\nu / \omega - \xi^2 p_\nu / \omega^2$ is a suitable projection vector of $F^{\nu}_{q_{ik}/q_i}$ and we have replaced the hadron $h$ by a parton $q_i$.

\sect{Result of matching coefficients}
As $K^{\nu}$ are independent of the hadronic state $\left | h \right \rangle$, we replace the hadronic state in Eq.~\eqref{eq:factorization} by a quark state and expand the formula perturbatively as
\begin{equation}
\begin{aligned} \label{eq:pfactorization}
&F^{\nu,(n)}_{q_{ik}/q_i} (\omega, \xi^2, \mu^2,\epsilon) =\\
&\sum_{m=0}^n \int_{-1}^1 \frac{dx}{x} f^{(m)}_{q_{ik}/q_i}(x,\mu^2,\epsilon) K^{\nu, (n-m)}(x\omega, \xi^2, \mu^2,\epsilon),
\end{aligned}
\end{equation}
with $n, m = 0, 1, 2, 3$ indicating the power of $\alpha_s$ and $f^{(n)}_{q_{ik}/q_i}$ being partonic PDF in the $\overline{\text{MS}}$ scheme\cite{Curci:1980uw, Moch:2004pa}. Note that, for partonic quantities, we use dimensional regularization with $d=4-2\epsilon$ to regularize both UV divergences and infrared/collinear divergences. Once $f^{(n)}_{q_{ik}/q_i}$ are known, we can derive the N3LO matching coefficient $K^{\nu (n)}$ by calculating $F^{\nu,(n)}_{q_{ik}/q_i}$ up to $n = 3$.

\begin{figure}[htbp]
\includegraphics[width=0.5\textwidth]{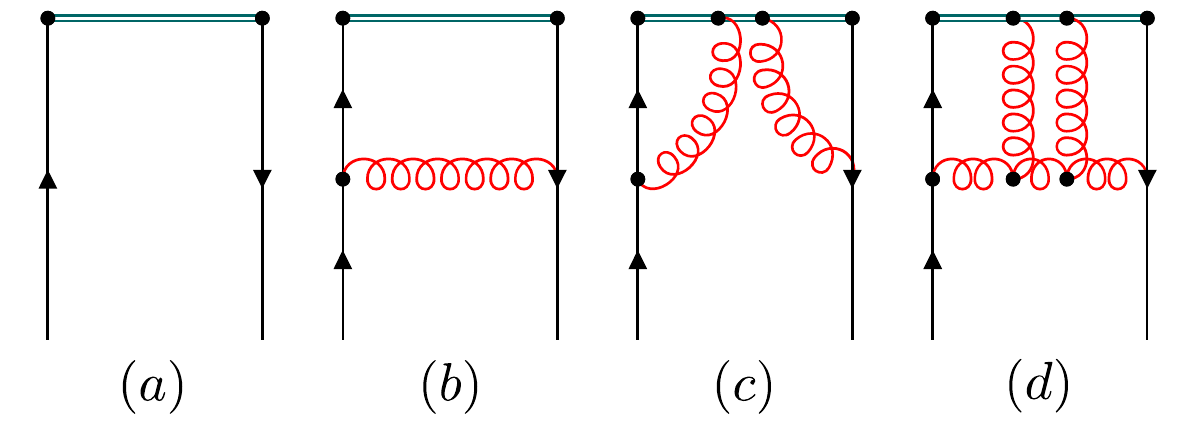}
\caption{\label{fig:feynmandiagrams}
Representative Feynman diagrams for QCFs at LO ($a$), NLO ($b$), NNLO ($c$) and N3LO ($d$). The double-line represents the gauge link.}
\end{figure}

In Fig.\ref{fig:feynmandiagrams}, we show some representative Feynman diagrams for $F^{\nu,(n)}_{q_{ik}/q_i}$. With $f^{(0)}_{q_{ik}/q_i}(x) = \delta(1-x)$, the diagram (a) gives the tree level result
\begin{align} \label{eq:LO}
    K^{\nu,(0)}(\omega) = F^{\nu,(0)}_{q_{ik}/q_i} (\omega)= -2 \mi p^\nu e^{i \omega}.
\end{align}
For diagrams at high orders, the amplitudes contain loop integrals and path-ordered integrals along the gauge link. For example, the diagram (c) gives
\begin{multline}
\begin{aligned} \label{eq:2loopamp}
M_c^\nu &= \frac{1}{2} g^4_s {C_F}^2 \mu^{4 \epsilon} \int_{0}^{1}dr_1 \int_{r_1}^{1}dr_2 \int \frac{d^d l_1 d^d l_2}{(2 \pi)^{2d}} \\
&\times \frac{e^{i r_1 l_2\cdot\xi + i r_2 l_1\cdot\xi + i (p-l_1)\cdot\xi}}{(l_1^2 + i 0^+)(l_2^2 + i 0^+)}\\
&\times \frac{\text{Tr}[\not{p}\not{\xi}(\not{p}-\not{l_1})\gamma^\nu (\not{p}+\not{l_2})\not{\xi}]}{((p-l_1)^2 + i 0^+)((p+l_2)^2 + i 0^+)}.
\end{aligned}
\end{multline}
As explained in Ref.\cite{Li:2020xml}, the exponential parts can be transformed to linear propagators in momentum space by using
\begin{multline}
\begin{aligned} \label{eq:fourier}
 \mathscr{F}&\left \lbrace \int_{0}^{1}dr_1 \int_{r_1}^{1}dr_2 \xi^2 e^{i r_2 l_1\cdot\xi + i r_1 l_2\cdot\xi + i (p-l_1)\cdot\xi} \right \rbrace \\
=&-\int d \xi_z e^{-i q_z \xi_z} \xi_z^2 \int_{0}^{1}dr_1 \int_{r_1}^{1}dr_2  \\
&\times e^{-i ((r_2-1) l_{1z} + r_1 l_{2z} p_z)\xi_z} \\
=& 2\text{Im}\frac{1}{(p_z + \tilde{q_z})(- l_{1z} + p_z + \tilde{q_z})(l_{2z} + p_z + \tilde{q_z})},
\end{aligned}
\end{multline}
where $\tilde{q_z} = q_z + i 0^+$ and $2 \pi \delta(x) = -2 \text{Im}(\frac{1}{x + i 0^+})$ has been used.

After performing the aforementioned transformation, loop integrals in momentum space are suitable to be handled using integration-by-parts relations \cite{Chetyrkin:1981qh, Laporta:2000dsw}, to express them as linear combinations of a small set of integrals, called master integrals (MIs). We use the package \texttt{Blade}\cite{Guan:2024byi,Guan:2019bcx} for reduction and \texttt{LiteRed+FiniteFlow} \cite{Lee:2013mka, Peraro:2019svx} to perform a crosscheck. For example, at the NLO we eventually have two MIs defined as:
\begin{align}
    I^{(1)}_1 &= \int \frac{d^d l_1}{(2 \pi)^{d}} \frac{1}{l_1^2 + i 0^+} 2 \text{Im} \frac{1}{l_{1z} + \tilde{q_z}},\\
    I^{(1)}_2 &= \int \frac{d^d l_1}{(2 \pi)^{d}} \frac{1}{l_1^2 + i 0^+} 2 \text{Im}\frac{1}{l_{1z} + p_z + \tilde{q_z}}.
\end{align}

Any MI generated from $F^{\nu,(n)}_{q_{ik}/q_i}$ can be expressed as 
\begin{align}
    I^{(n)}_j(p_z, q_z;\epsilon) = |q_z|^{(-1-2 n \epsilon)} q_z^{\delta^{(n)}_j} J^{(n)}_j (y;\epsilon),
\end{align}
where $y=p_z/q_z$ and $\delta^{(n)}_j-1-2 n \epsilon$ is the mass dimension of $I^{(n)}_j$. By applying integration-by-parts reduction, the differential equations of $J^{(n)}_j(y;\epsilon)$ can be constructed as \cite{Kotikov:1990kg}
\begin{align}
    \frac{\partial}{\partial y}\vec{J}^{(n)}(y;\epsilon) = A(y;\epsilon) \vec{J}^{(n)}(y;\epsilon).
\end{align}
The boundary conditions $J^{(n)}_j(0;\epsilon)$ are vacuum integrals with linear propagators at n-loop order, which can be calculated by the method of auxiliary mass flow\cite{Liu:2017jxz,Liu:2020kpc, Liu:2021wks, Liu:2022tji, Liu:2022mfb, Liu:2022chg}. Based on the differential equations and boundary conditions, $J^{(n)}_j(y;\epsilon)$ can be expanded as a Taylor series of $y$ to any needed order. Therefore, we can obtain results of QCFs as a Taylor series of $y$ by combining the coefficients of reduction and MIs.

After obtaining QCFs in momentum space, we transform the results to position space by inverse Fourier transformation, and the series of $y$ will be transformed to series of $\omega$. By adding contributions from all diagrams, applying the UV renormalization factor in the $\overline{\text{MS}}$ scheme and using the factorization formula \eqref{eq:pfactorization}, we obtain the matching coefficients $K^{\nu (n)}$ up to N3LO. All divergences are canceled and the results of $K^{\nu (n)}$ are finite after taking $\epsilon \rightarrow 0$, which verifies the proof of the factorization theorem\cite{Ma:2014jla}. Our NLO and NNLO results agree with previous calculations in Ref.\cite{Ma:2017pxb,Li:2020xml}, and the new results at N3LO are given in Appendix \ref{sec:app}.

\begin{figure}[htbp]
\begin{center}
     \includegraphics[width=3.5in]{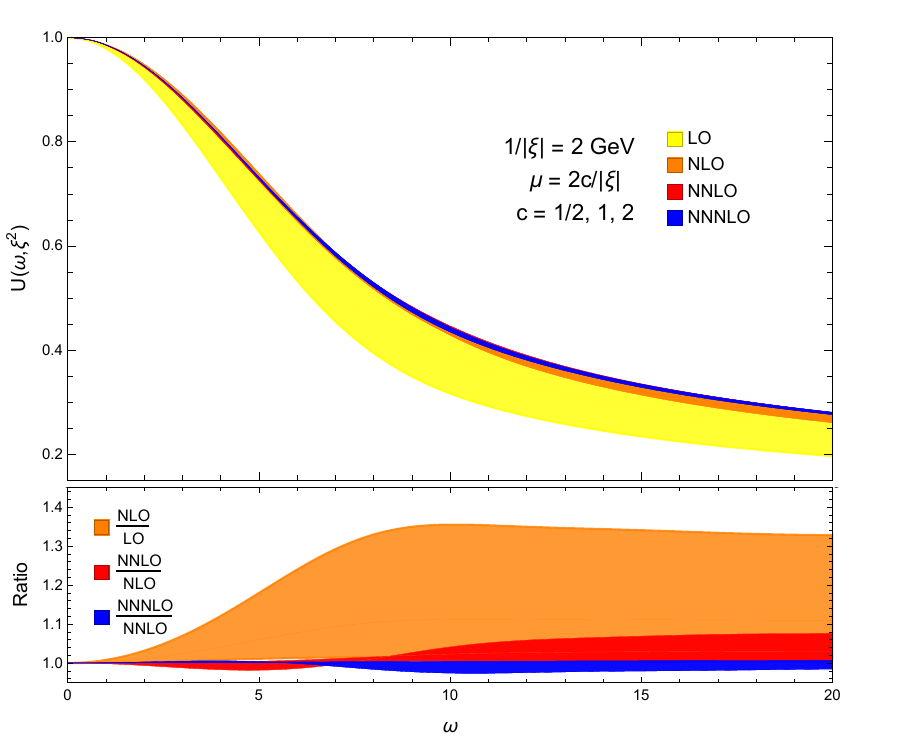}
	\includegraphics[width=3.55in]{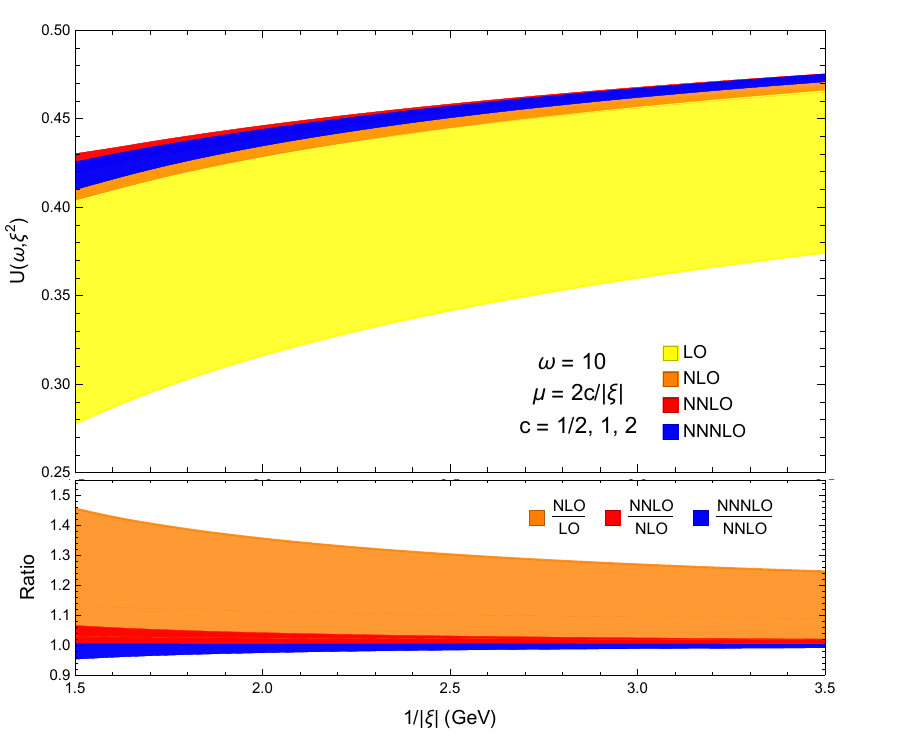}
\caption{\label{valenceplot}
Numerical predictions for valence-quark correlation functions with LO, NLO, NNLO and NNNLO matching coefficients and CT18NNLO PDFs.}
\end{center}
\end{figure}

By choosing $q_{k}$ as $\bar{q}_i$, we obtain valence-quark correlation functions. In this case, the factorization formula in Eq.~\eqref{eq:factorization} is modified as
\begin{equation}
\begin{aligned} \label{eq:factorization2}
&F^{\nu,\overline{\text{MS}}}_{q_{v}/h} (\omega, \xi^2, \mu^2) = \int_{0}^1 \frac{dx}{x} f_{q_{v}/h}(x,\mu^2) K_v^{\nu}(x\omega, \xi^2,\mu^2) , 
\end{aligned}
\end{equation}
where $q_v\equiv q_i-\bar{q}_i$, $F^{\nu,\overline{\text{MS}}}_{q_{v}/h} (\omega, \xi^2, \mu^2)\equiv F^{\nu,\overline{\text{MS}}}_{q_{i}/h} (\omega, \xi^2, \mu^2)-F^{\nu,\overline{\text{MS}}}_{\bar{q}_{i}/h} (\omega, \xi^2, \mu^2)$, $f_{q_{v}/h}(x,\mu^2)\equiv f_{q_{i}/h}(x,\mu^2)-f_{\bar{q}_{i}/h}(x,\mu^2)$ and $K_v^\nu\left(x \omega, \xi^2,\mu^2\right)\equiv K^{\nu}\left(x \omega,\xi^2, \mu^2\right)-K^{\nu}\left(-x \omega,\xi^2, \mu^2\right)$. With the obtained three-loop results of $K^{\nu}$, we can predict valence-quark correlation functions by using existing PDFs extracted from experimental data. In Fig. \ref{valenceplot}, we present $U\left(\omega,\xi^2\right)\equiv \frac{\mathrm{i}}{4\omega}\xi\cdot F_{q_v/h}\left(\omega,\xi^2\right)/\lim\limits_{\omega\to 0}\frac{\mathrm{i}}{4\omega}\xi\cdot F_{q_v/h}\left(\omega,\xi^2\right)$ either as a function of $\omega$ with a fixed $1/|\xi|=2 $ GeV, or as a function of $1/|\xi|$ with a fixed $\omega=10$. As the N3LO PDF is not available, we use CT18NNLO PDFs \cite{Hou:2019efy} as our input, and set $\mu=2c/|\xi|$ to minimize logarithms encountered in perturbative calculation. We vary $c=\frac{1}{2}, 1, 2$ for the bands to estimate theoretical uncertainties due to the ambiguity of scale choice. Our results show an improvement in perturbative uncertainty when N3LO matching coefficients are used.

\sect{Extraction of splitting functions}

We define plus or minus flavor non-singlet PDF as 
\begin{multline}
    \begin{aligned} \label{eq:non-siglet}
    f^{\pm}_{ns}(x, \mu^2,\epsilon) &\equiv (f_{q_{i}/h}(x, \mu^2,\epsilon)\pm f_{\bar{q}_{i}/h}(x, \mu^2,\epsilon)) \\ &- (f_{q_{k}/h}(x, \mu^2,\epsilon)\pm f_{\bar{q}_{k}/h}(x, \mu^2,\epsilon)).
\end{aligned}
\end{multline}
In the $\overline{\text{MS}}$ renormalization scheme, these quantities satisfy diagonalized DGLAP evolution equations \cite{Gribov:1972ri,Altarelli:1977zs,Dokshitzer:1977sg}:
\begin{equation} \label{eq:DGLAPns}
    \frac{d}{d\ \ln{\mu^{2}}} f^{\pm}_{ns}(x, \mu^{2},\epsilon)
        = P^{\pm}_{ns}(x, \mu^{2}) \otimes  f^{\pm}_{ns}(x, \mu^{2},\epsilon),
\end{equation}
where $P^{\pm}_{ns}$ are splitting functions and $\otimes$ represents the standard Mellin convolution 
\begin{equation}
    A(x) \otimes B(x) \equiv \int_{x}^{1} \frac{dz}{z} A(z) B(\frac{x}{z})\,.
\end{equation} 
The splitting functions can be perturbatively expanded as
\begin{equation} \label{eq:pexpand}
  P^{\pm}_{ns} (x, \mu^{2}) = \sum_{n=1}^\infty \alpha_s^{n} P^{\pm(n)}_{ns} (x), 
\end{equation}
and our goal is to extracted $P^{\pm(3)}_{ns} (x)$.

By inserting the solution of Eq.~\eqref{eq:DGLAPns} into
Eq.~\eqref{eq:pfactorization}, 
we can eventually express the divergences of ${F}^{\nu,(n)}_{q_{ik}/q_i}$ in the following form (with a detailed derivation given in Appendix \ref{sec:dglap}):
\begin{multline}\label{eq:expansion}
    \begin{aligned}
&-\int_0^1\mathrm{d}x [\cos(x\omega) P_{ns}^{-(n)}(x)+\mathrm{i} \sin(x\omega) P_{ns}^{+(n)}(x)]\\
&=\sum_{N=0}^{\infty} \frac{(-1)^N}{(2N)!} \gamma^{-(n)}_{ns} (2N+1) \ \omega^{2N} \\
&\quad + \sum_{N=0}^{\infty} \mathrm{i} \frac{(-1)^N}{(2N+1)!} \gamma^{+(n)}_{ns} (2N+2) \ \omega^{2N+1},
    \end{aligned}
\end{multline}
where $\gamma^{\pm(n)}_{ns} (N)$ is the Mellin transformation of $P^{\pm(n)}_{ns}(x)$. 
Therefore, with the series expansion of $\omega$ for the divergent part of ${F}^{\nu,(n)}_{q_{ik}/q_i}$, we can extract $\gamma^{\pm(n)}_{ns} (N)$ from each order of $\omega$ and thus get $P^{\pm(n)}_{ns}(x)$ by inverse Mellin transformation.

In the actual extraction of the splitting function at $n = 3$, we make ansatzs for the divergent part of ${F}^{\nu,(3)}_{q_{ik}/q_i}$ in terms of the basis 
\begin{multline}
    \begin{aligned}
    &\int_0^1 \mathrm{d}z\frac{p(z)H(\mathbf{n},z)}{1-z}\left(e^{\mathrm{i}z\omega}-e^{\mathrm{i\omega}}\right), \\
    &\int_{-1}^0 \mathrm{d}z\frac{p(z)H(\mathbf{n},-z)}{1-z}\left(e^{\mathrm{i}z\omega}-e^{\mathrm{i\omega}}\right),
    \end{aligned}
\end{multline}
where $H(\mathbf{n},z)$ are harmonic polylogorithms\cite{Remiddi:1999ew,Maitre:2005uu} with transcendentality equal or lower than 4, and $p(z)$ are polynomials of $z$ with degree equal or lower than 1. We expand the divergent part of ${F}^{\nu,(3)}_{q_{ik}/q_i}$ to 1000-order in $\omega$ to gain the solution of the ansatz, and obtain the analytic result with exact rational and irrational numbers by using the PSLQ algorithm\cite{1999MaCom..68..351F}. Then, using the series formula of $\omega$ for the analytic results, we obtain $\gamma^{(3)}_{ns} (N)$ as functions of $N$ and transform to $P^{(3)}_{ns}(x)$. Our results agree with the state-of-the-art result in Ref. \cite{Moch:2004pa}.

Compared with the traditional way of extracting splitting functions, where one makes use of deep-inelastic structure functions \cite{Moch:2004pa}, our method of using QCFs is much more advantageous. The advantages are already obvious at tree level. For QCFs, there is only one Feynman diagram, namely Fig. \ref{fig:feynmandiagrams}(a), which can be computed algebraically. However, for deep-inelastic structure functions, there are two Feynman diagrams. One can be obtained by replacing the gauge link in \ref{fig:feynmandiagrams}(a) with a quark line, and the other can be obtained by further reversing the orientation of quark lines. Besides having more Feynman diagrams, momentum will flow into the intermediate quark line and one needs to integrate over the momentum, which entails much more complex computations. These advantages have a huge impact at higher orders. Therefore,  compared with using deep-inelastic structure functions, it will be much more optimistic to extract four-loop splitting functions using QCFs.

\sect{Summary}
Quark correlation functions  (QCFs) are calculable in LQCD and factorizable to PDFs \cite{Ma:2017pxb, Ma:2014jla, Izubuchi:2018srq}, thus establishing connections between perturbative QCD and LQCD. For the first time, we calculate the complete N3LO unpolarized flavor non-singlet matching coefficients for QCFs in the $\overline{\text{MS}}$ renormalization scheme. With these N3LO coefficient functions and existing PDFs, we predict the valence-quark correlation function as shown in Fig.\ref{valenceplot}. This indicates a reduction in scale-variation uncertainties compared to NNLO results, demonstrating convergence as higher-order corrections are considered. Comparing our predictions with future LQCD results will enable the extraction of N3LO PDFs.

Furthermore, based on the divergent part of QCFs, we have extracted the flavor non-singlet splitting function up to three-loop, which is consistent with the results in the literature. This method presents a new means of calculating the splitting function. Thanks to the simplicity of QCFs, extracting the splitting function from QCFs has great potential to assist in achieving the calculation of the complete four-loop splitting function, which is extremely important at present.

\begin{acknowledgments}
We thank X. Guan and Z. Liu for many helpfull discussions. The work was supported in part by the National Natural Science Foundation of China (Grants No. 12325503, No. 11975029), the National Key Research and Development Program of China under	Contracts No.~2020YFA0406400. Z.Li was supported in part by the U.S. Department of Energy (DOE) Contract No.~DE-AC05-06OR23177.
We acknowledge the computational support from the High-performance Computing Platform of Peking University. Feynman diagrams are drawn using the {\tt FeynGame} program \cite{Harlander:2020cyh}.
\end{acknowledgments}

\appendix

\section{Relating QCFs to splitting functions}
\label{sec:dglap}
The evolution equations \eqref{eq:DGLAPns} hold both at the hadronic level and the partonic level. Partonic level non-singlet PDF can be formally expanded as:
\begin{equation} \label{eq:fexpand}
  f^{\pm}_{ns} (x, \mu^{2}, \epsilon) = \sum_{n=0}^\infty \alpha_s^{n} f^{\pm(n)}_{ns} (x, \epsilon).
\end{equation}
Note that $f^{\pm(n)}_{ns}$ and $P^{\pm(n)}_{ns}$ are independent of the scale $\mu$ and all scale dependences are taken care of by the $\alpha_s$, following the running coupling equations:
\begin{equation} \label{eq:running}
    \frac{d\ \alpha_s}{d\ \ln{\mu^{2}}} = -\epsilon \alpha_s + \sum_{n=0}^{\infty} \beta_n \alpha_s^{n+2},
\end{equation}
where $\beta_n$ are coefficients of the beta-function in QCD. By inserting Eqs.~\eqref{eq:pexpand}, \eqref{eq:fexpand} and \eqref{eq:running} into Eq.~\eqref{eq:DGLAPns}, we obtain relations by matching each power of $\alpha_s$ from both sides of the equations:
\begin{multline}
\begin{aligned} \label{eq:DGLAPexpand}
	-\epsilon n\ f^{\pm(n)}_{ns} (x, \epsilon) &= \sum_{m=0}^{n-1} P^{\pm(n-m)}_{ns}(x) \otimes f^{\pm(m)}_{ns} (x, \epsilon) \\
 &- \sum_{m=1}^{n-1} m \ \beta_{n-1-m}\ f^{\pm(m)}_{ns} (x, \epsilon).
\end{aligned}
\end{multline}

The definition of PDF is normalized so that $f^{\pm(0)}_{ns}(x, \epsilon) = \delta(1-x)$, and $f^{\pm(n)}_{ns} (x, \epsilon)$ with $n>0$ can be expanded as 
\begin{align} \label{eq:epsilonseries}
  f^{\pm(n)}_{ns} (x, \epsilon) &= \sum_{i=1}^n \frac{f^{\pm(n,i)}_{ns} (x)}{\epsilon^i}, 
\end{align}
where there is no finite terms in the expansion. With this information in hand, by matching finite terms on both sides of Eq.~\eqref{eq:DGLAPexpand} we get 
\begin{align}\label{eq:epsilonpole}
f^{\pm(n,1)}_{ns} (x) &= -\frac{P^{\pm(n)}_{ns}(x)}{n},
\end{align}
which expresses the $\epsilon^{-1}$ pole of the $n$-loop partonic PDF by the $n$-loop splitting function.
Furthermore, by matching $\epsilon^{-i}$ ($i\geq 1$) pole in Eq.~\eqref{eq:DGLAPexpand}, we get 
\begin{multline}
\begin{aligned} 
	f^{\pm(n,i+1)}_{ns} (x) &= -\frac{1}{n}\sum_{m=i}^{n-1} P^{\pm(n-m)}_{ns}(x) \otimes  f^{\pm(m,i)}_{ns} (x)  \\
 &+\frac{1}{n} \sum_{m=i}^{n-1} m \ \beta_{n-1-m}\ f^{\pm(m,i)}_{ns} (x),
\end{aligned}
\end{multline}
which eventually expresses higher-order poles of the $n$-loop partonic PDF by splitting functions with less than $n$ loops.

As our goal is to determine $n$-loop splitting functions, we assume that the splitting functions $P^{\pm(i)}_{ns}$ and the matching coefficients $K^{\nu (i)}$ are known for $i<n$. Thus we can rearrange Eq.~\eqref{eq:pfactorization} by moving known results to the left-hand-side of the equation,
\begin{equation}
\begin{aligned} \label{eq:pfactorization2}
&\tilde{F}^{\nu,(n)}_{q_{ik}/q_i} (\omega, \xi^2, \mu^2,\epsilon) =\text{finite terms}\\
&\quad+\frac{1}{\epsilon}\int_{-1}^1 \frac{dx}{x}  {f^{(n,1)}_{q_{ik}/q_i} (x)} \times K^{\nu (0)}(x\omega, \xi^2, \mu^2,\epsilon),
\end{aligned}
\end{equation}
where
\begin{equation}
    \begin{aligned}
        &\tilde{F}^{\nu,(n)}_{q_{ik}/q_i}(\omega, \xi^2, \mu^2,\epsilon) \equiv F^{\nu,(n)}_{q_{ik}/q_i} (\omega, \xi^2, \mu^2,\epsilon)\\
       &  -\sum_{m=1}^{n-1} \int_{-1}^1 \frac{\mathrm{d}x}{x} f^{(m)}_{q_{ik}/q_i}(x,,\epsilon) K^{\nu (n-m)}(x\omega, \xi^2, \mu^2,\epsilon)\\
       &  -\int_{-1}^1 \frac{dx}{x} \sum_{j=2}^n \frac{f^{(n,j)}_{q_{ik}/q_i} (x)}{\epsilon^j} \times K^{\nu (0)}(x\omega, \xi^2, \mu^2,\epsilon).
    \end{aligned}
\end{equation}
It provides the information that, by computing the divergences of ${F}^{\nu,(n)}_{q_{ik}/q_i}$,  we can extract $f^{(n,1)}_{q_{ik}/q_i} (x)$ and thus $P^{\pm(n)}_{ns}(x)$.

By contracting both sides of Eq.~\ref{eq:pfactorization2} by $\mi\frac{n}{2}( \xi_\nu / \omega - \xi^2 p_\nu / \omega^2)$, and making use of the relation $f_{q_i / h}(x) = -f_{\bar{q}_i / h}(-x)$ as well as the Eq.~\eqref{eq:LO}, the divergent part of Eq.~\ref{eq:pfactorization2} becomes 
\begin{multline}\label{eq:scalar}
    \begin{aligned}
&\mi\frac{n}{2}( \xi_\nu / \omega - \xi^2  p_\nu / \omega^2) \int_{-1}^1 \frac{\mathrm{d}x}{x}  {f^{(n,1)}_{q_{ik}/q_i} (x)} \times (-2 \mi x p^\nu e^{i x\omega})\\
&= n \int_{0}^1 \mathrm{d}x {f^{(n,1)}_{q_{ik}/q_i} (x)} e^{i x\omega} + {f^{(n,1)}_{q_{ik}/q_i} (-x)} e^{-i x\omega}\\
& = n \int_{0}^1 \mathrm{d}x \left(f^{(n,1)}_{q_{ik}/q_i} (x) - f^{(n,1)}_{\bar{q}_{ik}/q_i} (x)\right) \cos(x\omega) \\
& \qquad\qquad + i \left(f^{(n,1)}_{q_{ik}/q_i} (x) + f^{(n,1)}_{\bar{q}_{ik}/q_i} (x)\right) \sin(x\omega)\\
& = n \int_{0}^1 \mathrm{d}x f^{-(n,1)}_{ns} (x) \cos(x\omega) + i f^{+(n,1)}_{ns} (x)\sin(x\omega)\\
&=-\int_0^1\mathrm{d}x [\cos(x\omega) P_{ns}^{-(n)}(x)+\mathrm{i} \sin(x\omega) P_{ns}^{+(n)}(x)],
\end{aligned}
\end{multline}
which is desired form of Eq.~\eqref{eq:expansion}. 

\newpage
\begin{widetext}
\section{PERTURBATIVE RESULTS FOR QUARK CORRELATION FUNCTION}
\label{sec:app}
Renormalization factor in $\overline{\text{MS}}$ subtraction scheme is obtained as
\begin{multline}
    \begin{aligned}
Z^{\overline{\text{MS}}}=&1+\frac{\alpha_s S_\epsilon}{\pi\epsilon}C_F\frac{3}{4}+\left(\frac{\alpha_s S_\epsilon}{\pi\epsilon}\right)^2 C_F\left\lbrace\left[\frac{n_f}{16}+\frac{9C_F}{32} -\frac{11C_A}{32}\right]+\left[-\frac{5n_f}{96}+\left(\frac{\pi^2}{12}-\frac{5}{64}\right)C_F+\left(-\frac{\pi^2}{48}+\frac{49}{192}\right)C_A\right]\epsilon\right\rbrace\\
&+\left(\frac{\alpha_s S_\epsilon}{\pi\epsilon}\right)^3C_F\left\lbrace\left[\frac{n_f^2}{144}+n_f\left(\frac{3C_F}{64}-\frac{11C_A}{144}\right)+\frac{9C_F^2}{128}-\frac{33C_A C_F}{128}+\frac{121C_A^2}{576}\right]+\epsilon\left[-\frac{5n_f^2}{864}+C_F C_A\left(\frac{551}{2304}-\frac{115\pi^2}{1728}\right)\right.\right.\\
&\left.+C_A^2\left(-\frac{1151}{3456}+\frac{11\pi^2}{864}\right)+C_F^2\left(-\frac{15}{256}+\frac{\pi^2}{16}\right)+n_f C_A\left(\frac{97}{864}-\frac{\pi^2}{432}\right)+n_f C_F\left(-\frac{19}{1152}+\frac{\pi^2}{108}\right)\right]+\epsilon^2\left[-\frac{35n_f^2}{5184}\right.\\
&\left.+n_f C_A\left(\frac{4}{81}+\frac{7\pi^2}{1296}-\frac{19\zeta(3)}{144}\right)+n_f C_F\left(-\frac{235}{1728}-\frac{7\pi^2}{324}+\frac{11\zeta(3)}{72}\right)+C_A C_F\left(\frac{655}{6912}+\frac{37\pi^2}{324}+\frac{\pi^4}{1080}-\frac{71\zeta(3)}{288}\right)\right.\\
&\left.\left.+C_A^2\left(-\frac{1451}{20736}-\frac{65\pi^2}{2592}+\frac{\pi^4}{720}+\frac{11\zeta(3)}{288}\right)+C_F^2\left(\frac{37}{384}-\frac{\pi^2}{18}+\frac{\pi^4}{216}+\frac{3\zeta(3)}{16}\right)\right]\right\rbrace,
    \end{aligned}
\end{multline}
where $S_\epsilon\equiv\frac{(4\pi)^\epsilon}{\Gamma(1-\epsilon)}$ is a conventional factor in the $\overline{\text{MS}}$ scheme.

We express the matching coefficients $K^\nu\left(x \omega, \xi^2,\mu^2\right)\equiv x p^\nu A\left(x \omega,\xi^2,\mu^2\right)+x \omega\frac{\xi^\nu}{-\xi^2}B\left(x\omega,\xi^2,\mu^2\right)$, where $A(\omega, \xi^2, \mu^2)$ and $B(\omega, \xi^2, \mu^2)$ are analytical functions of $\omega$ everywhere except infinity. $A\left(x \omega,\xi^2,\mu^2\right)$ and $B\left(x \omega,\xi^2,\mu^2\right)$ can be perturbatively decomposed as 
\begin{multline}
    \begin{aligned}
A\left(\omega,\xi^2,\mu^2\right)=&-2\mathrm{i}\mathrm{e}^{\mathrm{i}\omega}+\frac{\alpha_s}{\pi}A^{(1)}\left(\omega,\xi^2,\mu^2\right)+\frac{\alpha_s^2}{\pi^2}A^{(2)}\left(\omega,\xi^2,\mu^2\right)+\frac{\alpha_s^3}{\pi^3}A^{(3)}\left(\omega,\xi^2,\mu^2\right),
\end{aligned}
\end{multline}
\begin{multline}
    \begin{aligned}
B\left(\omega,\xi^2,\mu^2\right)=&\frac{\alpha_s}{\pi}B^{(1)}\left(\omega,\xi^2,\mu^2\right)+\frac{\alpha_s^2}{\pi^2}B^{(2)}\left(\omega,\xi^2,\mu^2\right)+\frac{\alpha_s^3}{\pi^3}B^{(3)}\left(\omega,\xi^2,\mu^2\right).
\end{aligned}
\end{multline}

The analytical expressions of $A^{(1)}, A^{(2)}, B^{(1)}$ and $B^{(2)}$ can be found in \cite{Li:2020xml}, and results of $A^{(3)}\left(\omega,\xi^2,\mu^2\right)$ and $B^{(3)}\left(\omega,\xi^2,\mu^2\right)$ are given as a power series of $\omega$. Here we show the $\omega$ series up to $\omega^{5}$:

\begin{align}
&A^{(3)}\left(\omega,\xi^2,\mu^2\right)=\mathrm{i}\;\left\lbrace C_F^3\left[- \frac{1273}{768} - \frac{41\pi^{2}}{48} + \frac{577\pi^{4}}{4320} - \frac{31\zeta(3)}{24} + \pi^{2}\zeta(3) - \frac{25\zeta(5)}{12} - \left( \frac{271}{256} + \frac{\pi^{2}}{6} + \frac{\pi^{4}}{36} - \frac{15}{8}\zeta(3) \right)L -\right.\right.\nonumber\\
&\left.\left( \frac{15}{32} + \frac{\pi^{2}}{4} \right)L^{2} - \frac{9}{64}L^{3}\right]+C_F^2 C_A\left[-\frac{9505}{1296} + \frac{9451\pi^{2}}{7776} - \frac{4927\pi^{4}}{25920} + \frac{691\zeta(3)}{108} - \frac{5}{3}\pi^{2}\zeta(3) + \frac{49\zeta(5)}{4} + \right.\nonumber\\
&\left.\left( -\frac{2643}{256} - \frac{181\pi^{2}}{288} - \frac{\pi^{4}}{180} + \frac{129\zeta(3)}{16} \right)L + \left( -\frac{587}{192} - \frac{35\pi^{2}}{144} \right)L^{2} - \frac{33}{64}L^{3}\right]+C_F C_A^2\left[- \frac{2076127}{62208} + \frac{4523\pi^{2}}{31104} + \frac{533\pi^{4}}{12960} - \right.\nonumber\\
&\left.\frac{715\zeta(3)}{108} + \frac{97}{144}\pi^{2}\zeta(3) - \frac{461\zeta(5)}{96} + \left( - \frac{14579}{864} + \frac{115\pi^{2}}{216} - \frac{\pi^{4}}{120} - \frac{33\zeta(3)}{16} \right)L + \left( - \frac{685}{192} + \frac{11\pi^{2}}{144} \right)L^{2} - \frac{121}{288}L^{3}\right]+\nonumber\\
&C_F^2 n_f\left[\frac{48989}{10368} + \frac{61\pi^{2}}{972} + \frac{223\pi^{4}}{6480} - \frac{809\zeta(3)}{216} + \left( \frac{1151}{384} + \frac{\pi^{2}}{6} - \frac{9\zeta(3)}{4} \right)L + \left( \frac{2}{3} + \frac{\pi^{2}}{18} \right)L^{2} + \frac{3}{32}L^{3}\right]+C_F C_A n_f\left[\frac{181643}{15552} -\right.\nonumber\\
&\left.\frac{695\pi^{2}}{7776} - \frac{137\pi^{4}}{12960} + \frac{659\zeta(3)}{216} + \left( \frac{1303}{216} - \frac{11\pi^{2}}{108} + \frac{9\zeta(3)}{8} \right)L + \left( \frac{41}{32} - \frac{\pi^{2}}{72} \right)L^{2} + \frac{11}{72}L^{3}\right]+C_F n_f^2\left[- \frac{15373}{15552} - \frac{\zeta(3)}{12} - \right.\nonumber\\
&\left.\left.\frac{217L}{432} - \frac{5L^{2}}{48} - \frac{L^{3}}{72}\right]\right\rbrace+\left\lbrace C_F^3\left[-\frac{988169}{186624} + \frac{1561\pi^{2}}{3888} - \frac{2947\pi^{4}}{12960} + \frac{2561\zeta(3)}{216} + \frac{11}{9}\pi^{2}\zeta(3) - \frac{25\zeta(5)}{2} + \left( \frac{394741}{62208} - \frac{833\pi^{2}}{648} + \right.\right.\right.\nonumber\\
&\left.\left.\frac{\pi^{4}}{36} - \frac{77\zeta(3)}{24} \right)L + \left( -\frac{5219}{2592} + \frac{17\pi^{2}}{36} \right)L^{2} + \frac{4913}{5184}L^{3}\right]+C_F^2 C_A\left[-\frac{1009303}{34992} - \frac{629\pi^{2}}{1728} + \frac{623\pi^{4}}{5184} + \frac{8513\zeta(3)}{432} + \frac{25}{12}\pi^{2}\zeta(3) \right.\nonumber\\
&\left.\left.- \frac{319\zeta(5)}{24} + \left( \frac{538939}{20736} - \frac{2647\pi^{2}}{2592} + \frac{\pi^{4}}{180} - \frac{451\zeta(3)}{48} \right)L + \left( \frac{2639}{1728} + \frac{3\pi^{2}}{16} \right)L^{2} + \frac{3179}{1728}L^{3}\right]+C_F C_A^2\left[\frac{8150305}{559872} + \frac{485\pi^{2}}{7776} -\right.\right.\nonumber\\
&\left.\left.\frac{67\pi^{4}}{1620} + \frac{9551\zeta(3)}{864} - \frac{17}{16}\pi^{2}\zeta(3) + \frac{257\zeta(5)}{32} + \left( \frac{219967}{7776} - \frac{25\pi^{2}}{24} + \frac{\pi^{4}}{120} + \frac{131\zeta(3)}{48} \right)L + \left( \frac{14695}{5184} - \frac{11\pi^{2}}{144} \right)L^{2} + \frac{2057}{2592}L^{3}\right]+\right.\nonumber\\
&\left. C_F^2 n_f\left[\frac{240893}{279936} - \frac{7\pi^{2}}{54} - \frac{199\pi^{4}}{6480} - \frac{109\zeta(3)}{216} + \left( -\frac{59309}{10368} + \frac{5\pi^{2}}{81} + \frac{35\zeta(3)}{12} \right)L + \left( -\frac{11}{24} - \frac{\pi^{2}}{18} \right)L^{2} - \frac{289}{864}L^{3}\right]+\right.\nonumber\\
&\left.C_F C_A n_f\left[-\frac{850973}{139968} + \frac{7\pi^{2}}{108} + \frac{49\pi^{4}}{4320} - \frac{1363\zeta(3)}{432} + \left( -\frac{18047}{1944} + \frac{7\pi^{2}}{36} - \frac{43\zeta(3)}{24} \right)L + \left( -\frac{2497}{2592} + \frac{\pi^{2}}{72} \right)L^{2} - \frac{187}{648}L^{3}\right]+\right.\nonumber\\
&\left.C_F n_f^2\left[\frac{68773}{139968} + \frac{2921L}{3888} + \frac{79L^{2}}{1296} + \frac{17L^{3}}{648} + \frac{17\zeta(3)}{108}\right]\right\rbrace\omega+\mathrm{i}\;\left\lbrace C_F^3\left[-\frac{170992627}{29859840} + \frac{252433\pi^{2}}{311040} - \frac{34001\pi^{4}}{259200} + \frac{141263\zeta(3)}{17280} +\right.\right.\nonumber\\
&\left.\left.\frac{2}{3}\pi^{2}\zeta(3) - \frac{635\zeta(5)}{96} + L\left(\frac{28165079}{3981312} - \frac{29701\pi^{2}}{25920} + \frac{\pi^{4}}{72} - \frac{235\zeta(3)}{96}\right) + L^{2}\left(-\frac{566611}{165888} + \frac{43\pi^{2}}{144}\right) + \frac{79507L^{3}}{82944}\right]+\right.\nonumber\\
&\left.C_F^2 C_A\left[-\frac{1181654573}{35831808} - \frac{1489\pi^{2}}{46080} + \frac{18491\pi^{4}}{518400} + \frac{601403\zeta(3)}{34560} + \frac{929}{480}\pi^{2}\zeta(3) - \frac{399\zeta(5)}{32} + L\left(\frac{1890239}{82944} - \frac{110657\pi^{2}}{103680} + \frac{\pi^{4}}{360} -\right.\right.\right.\nonumber\\
&\left.\left.\left.\frac{839\zeta(3)}{192}\right) + L^{2}\left(-\frac{35473}{27648} + \frac{5\pi^{2}}{64}\right) + \frac{20339L^{3}}{13824}\right]+C_F C_A^2\left[-\frac{1440442991}{179159040} + \frac{657755\pi^{2}}{995328} - \frac{21683\pi^{4}}{1036800} + \frac{161513\zeta(3)}{17280} - \right.\right.\nonumber\\
&\left.\frac{2443\pi^{2}\zeta(3)}{2880} +\frac{2557\zeta(5)}{384} + L \left( \frac{9996893}{497664} - \frac{487\pi^{2}}{576} + \frac{\pi^{4}}{240} + \frac{253\zeta(3)}{192} \right)+ L^{2} \left( \frac{9043}{10368} - \frac{11\pi^{2}}{288} \right) + \frac{5203L^{3}}{10368} \right]+\nonumber\\
&\left.C_F^2n_f\left[\frac{240795887}{89579520} - \frac{1027\pi^{2}}{10368} - \frac{371\pi^{4}}{25920} - \frac{5131\zeta(3)}{4320}+ L \left( -\frac{1336223}{331776} + \frac{217\pi^{2}}{3240} + \frac{79\zeta(3)}{48} \right)+ L^{2} \left( \frac{2}{27} - \frac{\pi^{2}}{36} \right)- \frac{1849L^{3}}{6912}\right]\right.\nonumber\\
& C_F C_A n_f\left[\frac{19367557}{44789760} - \frac{587\pi^{2}}{23040} + \frac{257\pi^{4}}{43200} - \frac{3317\zeta(3)}{1728}+ L \left( -\frac{771949}{124416} + \frac{5\pi^{2}}{32} - \frac{13\zeta(3)}{12} \right)+ L^{2} \left( -\frac{6163}{20736} + \frac{\pi^{2}}{144} \right)- \frac{473L^{3}}{2592}\right]+\nonumber\\
&\left.C_F n_f^2\left[\frac{91159}{4478976} + \frac{29875 L}{62208} + \frac{127 L^2}{10368} + \frac{43 L^3}{2592} + \frac{43 \zeta(3)}{432}
\right] \right\rbrace\omega^2+\left\lbrace C_F^3\left[\frac{1116909401617}{279936000000} - \frac{120809533\pi^{2}}{116640000} + \frac{60631\pi^{4}}{1296000} - \right.\right.\nonumber\\
&\left.\left.\frac{1067011\zeta(3)}{1296000} + \frac{26\pi^{2}\zeta(3)}{135} - \frac{149\zeta(5)}{288}+ L \left( -\frac{11413637869}{2488320000} + \frac{1029289\pi^{2}}{1944000} - \frac{\pi^{4}}{216} + \frac{7091\zeta(3)}{7200}\right)+\right.\right.\nonumber\\
&\left.\left.L^{2} \left( \frac{678403531}{311040000} - \frac{247\pi^{2}}{2160} \right)- \frac{15069223L^{3}}{31104000}\right]+C_F^2 C_A\left[\frac{6129153542737}{335923200000} + \frac{165525563\pi^{2}}{466560000} - \frac{39691\pi^{4}}{7776000} - \frac{27343103\zeta(3)}{2592000} - \right.\right.\nonumber\\
&\left.\left.\frac{1111\pi^{2}\zeta(3)}{864} + \frac{833\zeta(5)}{96} + L\left(-\frac{587776447}{51840000} + \frac{829613\pi^{2}}{1555200} - \frac{\pi^{4}}{1080} + \frac{20527\zeta(3)}{14400}\right)+ L^{2}\left(\frac{12842437}{10368000} - \frac{193\pi^{2}}{8640}\right)\right.\right.\nonumber\\
&\left.\left.- \frac{671099L^{3}}{1036800} \right]+C_F C_A^2\left[\frac{105615680671}{13436928000} - \frac{69371549\pi^{2}}{124416000} + \frac{35057\pi^{4}}{5184000} - \frac{19337\zeta(3)}{5400} + \frac{3967\pi^{2}\zeta(3)}{8640} - \frac{4261\zeta(5)}{1152}+ \right.\right.\nonumber\\
&\left.\left.L\left(-\frac{175336129}{20736000} + \frac{3227\pi^{2}}{8640} - \frac{\pi^{4}}{720} - \frac{6389\zeta(3)}{14400}\right)+ L^{2}\left(-\frac{96379}{777600} + \frac{11\pi^{2}}{864}\right)- \frac{29887L^{3}}{155520}\right]+C_F^2 n_f\left[-\frac{299719154567}{167961600000} + \right.\right.\nonumber\\
&\left.\left.\frac{112561\pi^{2}}{2332800} + \frac{1759\pi^{4}}{388800} + \frac{8359\zeta(3)}{12960}+ L\left(\frac{1113393823}{622080000} - \frac{1537\pi^{2}}{48600} - \frac{427\zeta(3)}{720}\right)+ L^{2}\left(-\frac{32483}{216000} + \frac{\pi^{2}}{108}\right)+ \frac{61009L^{3}}{518400} \right]+\right.\nonumber\\
&\left. C_F C_A n_f\left[-\frac{4220484389}{3359232000} + \frac{1405793\pi^{2}}{46656000} - \frac{791\pi^{4}}{388800} + \frac{11953\zeta(3)}{16200}+ \left(\frac{39306187}{15552000} - \frac{11\pi^{2}}{160} + \frac{73\zeta(3)}{180}\right)L+ \left(\frac{65947}{1555200} - \frac{\pi^{2}}{432}\right)L^{2}+\right.\right.\nonumber\\
&\left.\left.\frac{2717L^{3}}{38880}  \right]+C_F n_f^2\left[\frac{106933397}{1679616000}- \frac{1487021L}{7776000}+ \frac{1049L^{2}}{777600}- \frac{247L^{3}}{38880}- \frac{247\zeta(3)}{6480}\right]\right\rbrace\omega^3+\mathrm{i}\left\lbrace C_F^3\left[\frac{19261627703441}{11757312000000} \right.\right.\nonumber\\
&\left.\left.- \frac{374944211\pi^{2}}{544320000} + \frac{15829\pi^{4}}{1296000} + \frac{265639\zeta(3)}{189000} + \frac{2701\pi^{2}\zeta(3)}{7560} - \frac{1243\zeta(5)}{576}+ L\left(-\frac{342604261}{186624000} + \frac{2233187\pi^{2}}{13608000} - \frac{\pi^{4}}{864} + \frac{2029\zeta(3)}{7200}\right)+\right.\right.\nonumber\\
&\left.\left.L^{2}\left(\frac{16316039}{19440000} - \frac{17\pi^{2}}{540}\right)- \frac{4913L^{3}}{30375}\right]+C_F^2 C_A\left[\frac{122985097981}{18662400000} + \frac{2282638583\pi^{2}}{6531840000} + \frac{181\pi^{4}}{1701000} - \frac{7190257\zeta(3)}{1512000} - \frac{5507\pi^{2}\zeta(3)}{8640} + \right.\right.\nonumber\\
&\left.\left.\frac{18055\zeta(5)}{4032}+ L\left(-\frac{398781841}{103680000} + \frac{1908209\pi^{2}}{10886400} - \frac{\pi^{4}}{4320} + \frac{1247\zeta(3)}{3600}\right)+ L^{2}\left(\frac{16255429}{31104000} - \frac{7\pi^{2}}{1440}\right)- \frac{3179L^{3}}{16200}\right]+\right.\nonumber\\
&\left.C_F C_A^2\left[\frac{45173470793}{13934592000} - \frac{137710511\pi^{2}}{522547200} + \frac{44377\pi^{4}}{27216000} - \frac{307747\zeta(3)}{403200} + \frac{1531\pi^{2}\zeta(3)}{8064} - \frac{8375\zeta(5)}{5376}+ L\left(-\frac{3754791157}{1492992000} + \frac{73\pi^{2}}{640} -\right.\right.\right.\nonumber\\
&\left.\left.\left.\frac{\pi^{4}}{2880} - \frac{1591\zeta(3)}{14400}\right)+ L^{2}\left(\frac{44707}{6220800} + \frac{11\pi^{2}}{3456}\right)- \frac{2057L^{3}}{38880}\right]+C_F^2 n_f\left[-\frac{132767455081}{195955200000} + \frac{5063\pi^{2}}{326592} + \frac{421\pi^{4}}{388800} + \frac{38387\zeta(3)}{181440}+ \right.\right.\nonumber\\
&\left.\left.L\left(\frac{87930883}{155520000} - \frac{3271\pi^{2}}{340200} - \frac{113\zeta(3)}{720}\right)+ L^{2}\left(-\frac{1107599}{15552000} + \frac{\pi^{2}}{432}\right)+ \frac{289L^{3}}{8100}\right]+C_F C_A n_f\left[-\frac{7873777739}{13436928000} + \frac{4399019\pi^{2}}{326592000} \right.\right.\nonumber\\
&\left.\left.- \frac{89\pi^{4}}{172800} + \frac{107507\zeta(3)}{518400}+ L\left(\frac{91729579}{124416000} - \frac{181\pi^{2}}{8640} + \frac{317\zeta(3)}{2880}\right)+ L^{2}\left(-\frac{469}{194400} - \frac{\pi^{2}}{1728}\right)+ \frac{187L^{3}}{9720}\right]+\right.\nonumber\\
&\left.C_F n_f^2\left[\frac{36559283}{1119744000} - \frac{17 \zeta(3)}{1620} - \frac{2563669 L}{46656000} + \frac{2437 L^{2}}{1555200} - \frac{17 L^{3}}{9720}\right]\right\rbrace\omega^4+\left\lbrace C_F^3\left[- \frac{493280354356168679}{988025713920000000} + \frac{1815838324799\pi^{2}}{5601052800000} - \right.\right.\nonumber\\
&\left.\left.\frac{2400641\pi^{4}}{952560000} - \frac{679431217\zeta(3)}{635040000} - \frac{8189\pi^{2}\zeta(3)}{37800} + \frac{443\zeta(5)}{320} + \left( \frac{207841327404241}{392073696000000} - \frac{128617007\pi^{2}}{3333960000} + \frac{\pi^{4}}{4320} - \frac{109357\zeta(3)}{1764000} \right)L + \right.\right.\nonumber\\
&\left.\left.\left( - \frac{420850033}{1823259375} + \frac{32\pi^{2}}{4725} \right)L^{2} + \frac{2097152}{52093125}L^{3} \right]+C_F^2 C_A\left[- \frac{16457144126742023}{9409768704000000} - \frac{4308726154297\pi^{2}}{22404211200000} - \frac{162761\pi^{4}}{635040000} + \right.\right.\nonumber\\
&\left.\left.\frac{208606297\zeta(3)}{118540800} + \frac{5027\pi^{2}\zeta(3)}{20160} - \frac{36409\zeta(5)}{20160}+ \left( \frac{1217627113207}{1244678400000} - \frac{16415447\pi^{2}}{381024000} + \frac{\pi^{4}}{21600} - \frac{59951\zeta(3)}{882000} \right)L+ \right.\right.\nonumber\\
&\left.\left.\left( - \frac{7868853211}{53343360000} + \frac{43\pi^{2}}{50400} \right)L^{2}+ \frac{22528}{496125}L^{3} \right]+C_F C_A^2\left[- \frac{963034914211721}{1075402137600000} + \frac{1361059801823\pi^{2}}{14936140800000} - \frac{238319\pi^{4}}{762048000} +\right.\right.\nonumber\\
&\left.\left.\frac{331517959\zeta(3)}{8890560000} - \frac{2579\pi^{2}\zeta(3)}{40320} + \frac{14299\zeta(5)}{26880} + \left( \frac{1477034490499}{2560481280000} - \frac{5347\pi^{2}}{201600} + \frac{\pi^{4}}{14400} + \frac{78103\zeta(3)}{3528000} \right)L+ \right.\right.\nonumber\\
&\left.\left.\left( - \frac{12865267}{1524096000} - \frac{11\pi^{2}}{17280} \right)L^{2}+ \frac{484}{42525}L^{3} \right]+C_F^2 n_f\left[\frac{5348255724227}{29042496000000} - \frac{21633823\pi^{2}}{5715360000} - \frac{2839\pi^{4}}{13608000} - \frac{811529\zeta(3)}{15876000} + \right.\right.\nonumber\\
&\left.L \left( -\frac{257081937097}{1867017600000} + \frac{105421\pi^{2}}{47628000} + \frac{827\zeta(3)}{25200} \right)+ L^{2} \left( \frac{112650109}{5334336000} - \frac{\pi^{2}}{2160} \right) - \frac{4096L^{3}}{496125} \right]+\nonumber\\
&\left.C_F C_A n_f\left[\left( \frac{136779268045867}{806551603200000} - \frac{1247383313\pi^{2}}{320060160000} + \frac{313\pi^{4}}{3024000} - \frac{11585683\zeta(3)}{254016000} \right) + L\left( -\frac{11874966559}{71124480000} + \frac{1471\pi^{2}}{302400} - \frac{2363\zeta(3)}{100800} \right) + \right.\right.\nonumber\\
&\left.\left.L^{2}\left( \frac{54883}{19051200} + \frac{\pi^{2}}{8640} \right) - \frac{176L^{3}}{42525} \right]+C_F n_f^2\left[-\frac{654261740623}{67212633600000} + \frac{32\zeta(3)}{14175} + \left(\frac{985741243}{80015040000}\right)L - \frac{205213}{381024000}L^2 +\right.\right.\nonumber\\
&\left.\left.\frac{16}{42525}L^3\right]\right\rbrace\omega^5,\nonumber\\
\end{align}
\begin{align}
&B^{(3)}\left(\omega,\xi^2,\mu^2\right)=\mathrm{i}\;\left\lbrace C_F^3\left[-\left(\frac{11}{128} + \frac{79\pi^2}{36}\right) + 10\zeta(3) + \frac{16\pi^2\zeta(3)}{9} - \frac{35\zeta(5)}{3} + L\left(-\frac{11}{16} + \frac{\pi^2}{3}\right) + \frac{9L^2}{32}\right]+\right.\nonumber\\
&\left.C_F^2C_A\left[\left(\frac{9443}{3456} + \frac{3013\pi^{2}}{1296} - \frac{7\pi^{4}}{540} - \frac{179\zeta(3)}{12} - \frac{13\pi^{2}\zeta(3)}{9} + \frac{65\zeta(5)}{6}\right) + L\left(\frac{5}{36} + \frac{13\pi^{2}}{54}\right) +\frac{33}{32}L^{2}\right]+\right.\nonumber\\
&\left.C_F C_A^2\left[\frac{10229}{2592} - \frac{269\pi^2}{288} - \frac{17\pi^4}{540} + \frac{161\zeta(3)}{24} + \frac{13\pi^2\zeta(3)}{36} - \frac{65\zeta(5)}{24} + L\left(\frac{1285}{432} - \frac{11\pi^2}{36}\right) + \frac{121L^2}{144}\right]+\right.\nonumber\\
&\left.C_F^2 n_f\left[-\left(\frac{2077}{1728} + \frac{7\pi^2}{81}\right) + \frac{3\zeta(3)}{2}+ L\left(-\frac{1}{72} - \frac{2\pi^2}{27}\right)-\frac{3L^2}{16}\right]+C_F C_An_f\left[-\frac{245}{324} + \frac{\pi^2}{24} + \frac{7\pi^4}{540} - \frac{17\zeta(3)}{12} + \right.\right.\nonumber\\
&\left.\left.L\left(-\frac{145}{216} + \frac{\pi^2}{18}\right) - \frac{11L^2}{36}\right]+C_F n_f^2\left[\frac{23}{648} + \frac{L}{108} + \frac{L^{2}}{36}
\right]\right\rbrace+\left\lbrace C_F^3\left[-\frac{132473}{31104} + \frac{524\pi^2}{81} - \frac{19\pi^4}{810} - \frac{463\zeta(3)}{18} - \frac{44\pi^2\zeta(3)}{9} +\right.\right.\nonumber\\
&\left.\frac{385\zeta(5)}{12} + L\left(\frac{3299}{1296} - \frac{7\pi^2}{24}\right) - \frac{289L^2}{864}\right]+C_F^2C_A\left[-\left(\frac{11237}{10368} + \frac{10171\pi^{2}}{1728} - \frac{139\pi^{4}}{6480} - \frac{1463\zeta(3)}{48} - \frac{53}{12}\pi^{2}\zeta(3) + \frac{785\zeta(5)}{24}\right)+ \right.\nonumber\\
&\left.\left. L\left(\frac{283}{216} - \frac{29\pi^{2}}{216}\right)-\frac{187L^{2}}{288}\right]+C_F C_A^2\left[-\frac{29947}{31104} + \frac{5183\pi^2}{3456} + \frac{103\pi^4}{4320} - \frac{2903\zeta(3)}{288} - \frac{10\pi^2\zeta(3)}{9} + \frac{415\zeta(5)}{48} +\right.\right.\nonumber\\
&\left.\left.L\left(-\frac{2561}{1296} + \frac{11\pi^2}{54}\right) - \frac{121L^2}{432}\right]+C_F^2 n_f\left[\frac{4225}{5184} + \frac{5\pi^{2}}{81} - \frac{35\zeta(3)}{36}+ L\left(-\frac{3}{8} + \frac{\pi^{2}}{18}\right)+\frac{17L^{2}}{144}\right]+\right.\nonumber\\
&\left.C_F C_An_f\left[\frac{959}{7776} + \frac{11L^2}{108} - \frac{5\pi^2}{2592} - \frac{7\pi^4}{720} + L\left(\frac{305}{648} - \frac{\pi^2}{27}\right) + \frac{125\zeta(3)}{144}
\right]+C_F n_f^2\left[-\frac{19}{1944} - \frac{5L}{324} - \frac{L^2}{108}
\right]\right\rbrace\omega+\nonumber\\
&\mathrm{i}\;\left\lbrace C_F^3\left[-\frac{33335689}{9953280} - \frac{1849L^2}{13824} + \frac{15787\pi^2}{2160} - \frac{211\pi^4}{16200} + L\left(\frac{59113}{41472} - \frac{2\pi^2}{15}\right) - \frac{3793\zeta(3)}{120} - \frac{52\pi^2\zeta(3)}{9} + \frac{455\zeta(5)}{12}\right]+\right.\nonumber\\
&\left.C_F^2C_A\left[-\left(\frac{42071}{41472} + \frac{108649\pi^{2}}{17280} - \frac{173\pi^{4}}{16200} - \frac{1347\zeta(3)}{40} - \frac{101}{20}\pi^{2}\zeta(3) + \frac{901\zeta(5)}{24}\right)+ L\left(\frac{2075}{3456} - \frac{127\pi^{2}}{2880}\right)-\frac{473L^{2}}{2304}\right]+\right.\nonumber\\
&\left.C_F C_A^2\left[\frac{1160077}{2488320} - \frac{121L^2}{1728} + \frac{20285\pi^2}{13824} + \frac{17\pi^4}{1800} + L\left(-\frac{3881}{5184} + \frac{11\pi^2}{144}\right) - \frac{9179\zeta(3)}{960} - \frac{877\pi^2\zeta(3)}{720} + \frac{925\zeta(5)}{96}
\right]+\right.\nonumber\\
&\left.C_F^2 n_f\left[\frac{352129}{829440} + \frac{13\pi^{2}}{1296} - \frac{133\zeta(3)}{360} + L\left(-\frac{83}{432} + \frac{\pi^{2}}{45}\right) + \frac{43L^{2}}{1152}\right]+C_F C_An_f\left[-\frac{12941}{155520} + \frac{11L^2}{432} + \frac{31\pi^2}{2880} - \frac{7\pi^4}{1800} + \right.\right.\nonumber\\
&\left.\left.L\left(\frac{883}{5184} - \frac{\pi^2}{72}\right) + \frac{\zeta(3)}{3}\right]+C_F n_f^2\left[-\frac{97}{31104} - \frac{13L}{2592} - \frac{L^2}{432}\right]\right\rbrace\omega^2+\left\lbrace C_F^3\left[\left( \frac{377099221}{248832000} - \frac{10072403\pi^{2}}{1944000} + \frac{67\pi^{4}}{16200} +\right.\right.\right.\nonumber\\
&\left.\left.\left.\frac{250583\zeta(3)}{10800} + \frac{568\pi^{2}\zeta(3)}{135} - \frac{497\zeta(5)}{18} \right) + \left( -\frac{671351}{1440000} + \frac{53\pi^{2}}{1296} \right)L + \frac{61009L^{2}}{1728000}
\right]+C_F^2C_A\left[\frac{13435787}{38880000} + \frac{24762463\pi^{2}}{5832000} - \right.\right.\nonumber\\
&\left.\left.\frac{47\pi^{4}}{14400} - \frac{74023\zeta(3)}{3240} - \frac{941}{270}\pi^{2}\zeta(3) + \frac{3755\zeta(5)}{144}+ L\left(-\frac{141331}{864000} + \frac{673\pi^{2}}{64800}\right)+\frac{2717L^{2}}{57600}\right]+C_F C_A^2\left[- \frac{18364519}{62208000} - \right.\right.\nonumber\\
&\left.\left.\frac{8881561\pi^{2}}{9331200} - \frac{677\pi^{4}}{259200} + \frac{1532381\zeta(3)}{259200} + \frac{1729\pi^{2}\zeta(3)}{2160} - \frac{923\zeta(5)}{144} + \left( \frac{719}{3600} - \frac{11\pi^{2}}{540} \right)L + \frac{121L^{2}}{8640}
\right]+\right.\nonumber\\
&\left.C_F^2 n_f\left[-\frac{41825447}{311040000} + \frac{\zeta(3)}{10} - \frac{73\pi^{2}}{194400} + L\left(\frac{37769}{648000}\right) - L\left(\frac{\pi^{2}}{162}\right)-\frac{247L^{2}}{28800}\right]+C_F C_An_f\left[\frac{158257}{3456000} - \frac{3757\pi^{2}}{777600} +\right.\right.\nonumber\\
&\left.\left.\frac{7\pi^{4}}{6480} - \frac{401\zeta(3)}{4320}+ L\left(-\frac{5681}{129600} + \frac{\pi^{2}}{270}\right)-\frac{11L^{2}}{2160}\right]+C_F n_f^2\left[\frac{2849}{3888000} + \frac{73L}{64800} + \frac{L^2}{2160}\right]\right\rbrace\omega^3+\nonumber\\
&\mathrm{i}\;\left\lbrace C_F^3\left[\frac{789893129}{1451520000} - \frac{48997507\pi^{2}}{18144000} + \frac{31\pi^{4}}{32400} + \frac{1858837\zeta(3)}{151200} + \frac{2099}{945}\pi^{2}\zeta(3) - \frac{2099\zeta(5)}{144}+ L\left(-\frac{8586239}{77760000} + \frac{143\pi^{2}}{15120}\right)+\right.\right.\nonumber\\
&\left.\left.\frac{289L^{2}}{40500}\right]+C_F^2C_A\left[\frac{140427581}{1866240000} + \frac{137185367\pi^{2}}{65318400} - \frac{673\pi^{4}}{907200} - \frac{5156533\zeta(3)}{453600} - \frac{139}{80}\pi^{2}\zeta(3) + \frac{13199\zeta(5)}{1008} + \right.\right.\nonumber\\
&\left.\left.\left(-\frac{171761}{5184000} + \frac{347\pi^{2}}{181440}\right)L + \frac{187}{21600}L^{2}\right]+C_F C_A^2\left[-\frac{15918143}{163296000} - \frac{73754449\pi^{2}}{163296000} - \frac{337\pi^{4}}{604800} + \frac{3316379\zeta(3)}{1209600} + \right.\right.\nonumber\\
&\left.\left.\frac{5731\pi^{2}\zeta(3)}{15120} - \frac{687\zeta(5)}{224}+ L\left(\frac{43037}{1036800} - \frac{11\pi^{2}}{2592}\right)+\frac{121L^{2}}{51840}\right]+C_F^2 n_f\left[-\frac{207885637}{6531840000} + \frac{149\pi^{2}}{453600} + \frac{319\zeta(3)}{15120}+ \right.\right.\nonumber\\
&\left.\left.L\left(\frac{33761}{2592000} - \frac{\pi^{2}}{756}\right)-\frac{17L^{2}}{10800}\right]+C_F C_An_f\left[\frac{2537543}{186624000} - \frac{14447\pi^{2}}{10886400} + \frac{\pi^{4}}{4320} - \frac{35\zeta(3)}{1728} + \left( -\frac{511}{57600} + \frac{\pi^{2}}{1296} \right)L - \right.\right.\nonumber\\
&\left.\left.\frac{11}{12960}L^{2}\right]+C_F n_f^2\left[\frac{791}{5832000} + \frac{13L}{64800} + \frac{L^2}{12960}\right]\right\rbrace\omega^4+\left\lbrace C_F^3\left[-\frac{13535060489789}{78414739200000} + \frac{178859789827\pi^{2}}{160030080000} - \frac{3349\pi^{4}}{19051200} -\right.\right.\nonumber\\
&\left.\left.\frac{651181997\zeta(3)}{127008000} - \frac{874}{945}\pi^{2}\zeta(3) + \frac{437\zeta(5)}{72}+ L\left(\frac{430310953}{20744640000} - \frac{19\pi^{2}}{10800}\right) -\frac{4096L^{2}}{3472875}\right]+C_F^2C_A\left[-\frac{1176425086637}{89616844800000} - \right.\right.\nonumber\\
&\left.\left.\frac{11728559497\pi^{2}}{14224896000} + \frac{10277\pi^{4}}{76204800} + \frac{126816289\zeta(3)}{28224000} + \frac{739\pi^{2}\zeta(3)}{1080} - \frac{3491\zeta(5)}{672} + \left(\frac{72282607}{13335840000} - \frac{17\pi^{2}}{58800}\right)L - \frac{44}{33075}L^{2} \right]+\right.\nonumber\\
&\left.C_F C_A^2\left[\frac{14309899067}{568995840000} + \frac{43365197237\pi^{2}}{256048128000} + \frac{4939\pi^{4}}{50803200}- \frac{258982249\zeta(3)}{254016000} - \frac{17}{120}\pi^{2}\zeta(3) + \frac{1555\zeta(5)}{1344}+\right.\right.\nonumber\\
&\left.\left.L\left(-\frac{150413}{21168000} + \frac{11\pi^{2}}{15120}\right)-\frac{121L^{2}}{362880}\right]+C_F^2 n_f\left[\frac{4467721529}{746807040000} - \frac{2827\pi^{2}}{25401600} - \frac{221\zeta(3)}{60480}+ L\left(-\frac{155783}{66679200} + \frac{\pi^{2}}{4320}\right)+\right.\right.\nonumber\\
&\left.\left.\frac{8L^{2}}{33075}\right]+C_F C_An_f\left[-\frac{103431133}{35562240000} + \frac{274201\pi^{2}}{1016064000} - \frac{7\pi^{4}}{172800} + \frac{2929\zeta(3)}{806400} + \left(\frac{47249}{31752000} - \frac{\pi^{2}}{7560}\right)L + \frac{11}{90720}L^{2}\right]+\right.\nonumber\\
&\left.C_F n_f^2\left[-\frac{1667}{80015040} - \frac{19L}{635040} - \frac{L^2}{90720}\right]\right\rbrace\omega^5,
\end{align}
where $L\equiv \ln\left(-\xi^2 \mu^2/4\right)+2\gamma_E$. Numerical results of $A^{(3)}\left(\omega,\xi^2,\mu^2\right)$ and $B^{(3)}\left(\omega,\xi^2,\mu^2\right)$ 
are provided in the Supplemental Material, which are sufficient to compute the matching coefficients $K^\nu$ up to $\left|\omega\right|=20$ with at least nine significant digits.

\end{widetext}

\providecommand{\href}[2]{#2}\begingroup\raggedright\endgroup

\end{document}